\documentclass[journal,compsoc]{IEEEtran}

\usepackage{booktabs} 
\usepackage{multirow} 
\usepackage{cite}
\usepackage{breakcites}
\usepackage{balance}
\usepackage{makecell}
\usepackage{comment}    
\usepackage[table]{xcolor}

\ifCLASSINFOpdf
\usepackage[pdftex]{graphicx}
\usepackage{caption}
\else
\fi

\begin{document}
\title{A Survey of Phase Classification Techniques for Characterizing Variable Application Behavior}

\author{Keeley~Criswell
	and~Tosiron~Adegbija,~\IEEEmembership{Member,~IEEE}
	
	\thanks{The authors are with the Department
		of Electrical and Computer Engineering, University of Arizona, USA, e-mail: \{kcriswell, tosiron\}@email.arizona.edu.
		
		The work was done while Keeley Criswell was with the University of Arizona.}
	}

\IEEEtitleabstractindextext{
\begin{abstract}
Adaptable computing is an increasingly important paradigm that specializes system resources to variable application requirements, environmental conditions, or user requirements. Adapting computing resources to variable application requirements (or application phases) is otherwise known as phase-based optimization. Phase-based optimization takes advantage of application phases, or execution intervals of an application that behave similarly, to enable effective and beneficial adaptability. In order for phase-based optimization to be effective, the phases must first be classified to determine when application phases begin and end, and ensure that system resources are accurately specialized. In this paper, we present a survey of phase classification techniques that have been proposed to exploit the advantages of adaptable computing through phase-based optimization. We focus on recent techniques, and classify these techniques with respect to several factors in order to highlight their similarities and differences. We divide the techniques by their major defining characteristics---online/offline and serial/parallel. In addition, we discuss other characteristics such as prediction and detection techniques, the characteristics used for prediction, interval type, etc. We also identify gaps in the state-of-the-art and discuss future research directions to enable and fully exploit the benefits of adaptable computing.

\end{abstract}

}

\maketitle

\section{Introduction}
\label{intro}

Much prior research has shown that applications have variable resource requirements throughout execution. Thus, for optimal execution, system resources (e.g., memory resources, clock frequency, functional units, etc.) must adapt to changes in application resource requirements. To enable this adaptability, \textit{application phases} \cite{denning68} \cite{sherwood99} specify execution intervals---typically measured by number of instructions executed or time periods---that exhibit similar execution behavior. Since a phase typically exhibits stable execution characteristics, the resource requirements for that phase are also stable \cite{balasubramonian00}  \cite{gordon08} \cite{kumar03} \cite{sherwood01}   \cite{srinivasan16}. \textit{Phase classification} groups intervals with similar characteristics to form phases, and represents a vital first step in adaptable computing \cite{gu08} \cite{khan10} \cite{khan11} \cite{peleg07} \cite{rodrigues13} \cite{shen07} \cite{srinivasan16} \cite{zhang15}. Phase classification offers several benefits for adaptable computing in the form of effective configuration of system components (at design time or during runtime), scheduling in heterogeneous systems, design-time rapid system evaluation and simulation, etc. 

\begin{figure}[t]
\centering
\includegraphics[width=.4\textwidth]{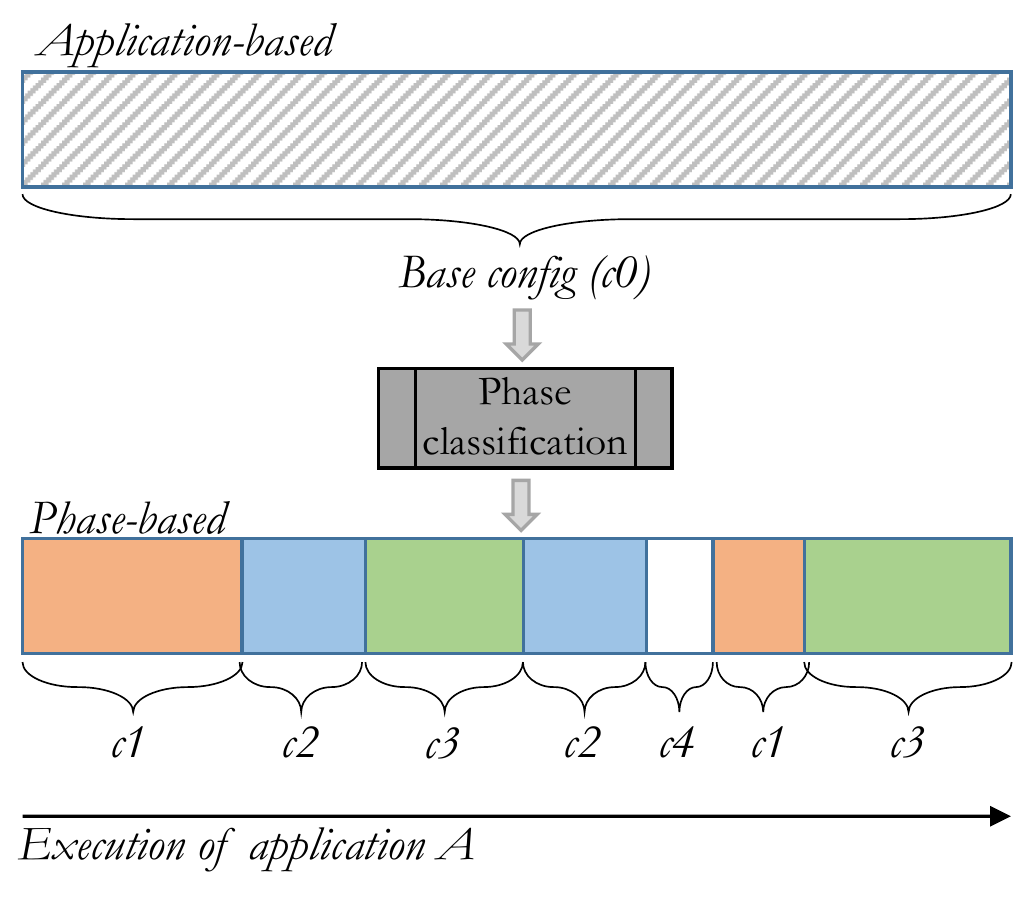}
\caption{Illustration of phase-based execution}
\label{fig:benefits}
\vspace{-5pt}
\end{figure}

Fig. \ref{fig:benefits} illustrates phase classification in the context of an application's execution. Rather than using a single configuration of system resources throughout the application's execution, phase classification determines the different application phases, such that each phase can be executed using the system configuration that most closely meets the phase's resource requirements. Prior work has shown that adapting system resources to application phases---\textit{phase-based optimization}---enables much higher optimization potential than application-based optimization \cite{adegbija12} \cite{gordon08} \cite{gu08} \cite{kumar03} \cite{kumar04} \cite{ratanaworabhan08} \cite{shen04} \cite{shen07} \cite{sherwood02} \cite{Sherwood03}. Phase-based optimization evaluates an application's characteristics and determines the best system configurations that meet each phase's execution characteristics. For example, Gordon-Ross et al. \cite{gordon08} found that phase-based cache optimization could yield 37\% and 20\% improvements in performance and energy usage, respectively, compared to application-based execution. 

Similarly, several other optimizations depend on phase classification \cite{dhodapkar02} \cite{kumar03} \cite{kumar04} \cite{shen07} \cite{peleg07} \cite{khan10} \cite{khan11} \cite{rodrigues13} \cite{srinivasan16}. For example, Khan et al. \cite{khan10} used phase classification to facilitate thread-to-core assignment in asymmetric multicore systems, in order to optimize throughput, power, or performance per watt. Dhodapkar et al. \cite{dhodapkar02} used adaptable instruction caches to meet each phase's instruction execution needs for optimized performance efficiency. 

Other optimizations, such as just-in-time compilers \cite{arnold00} \cite{arnold02}, dynamic instruction optimizations (e.g., \cite{bala00} \cite{merten01}), and performance bottleneck analysis \cite{georges04} also rely on accurate phase classification. Furthermore, phase classification enables rapid system evaluation and shorter simulation time during research. Rather than completely simulating an application, only a few phases that represent the full application's execution characteristics are simulated \cite{sherwood01} \cite{sherwood02}.  

Since phase classification is so important in the analysis and optimization of modern computing systems and applications, this paper focuses on surveying recent advances in phase classification. Several phase classification techniques have been proposed that cater to different system scenarios and tradeoffs. Phase classification techniques are usually categorized by the application characteristics used for classifying the phases. For example, Sherwood et al. \cite{sherwood01} found that an application's behavior is directly linked to the application code. The authors used the frequency of basic block execution---called \textit{basic block vectors (BBVs)}---to classify an application's phases. A basic block is a section of code with one entry and one exit that is executed from start to finish. 

Dhodapkar et al. \cite{dhodapkar03} compared three phase classification techniques that use instruction working sets  \cite{dhodapkar02}, basic block vectors (BBVs) \cite{sherwood02} \cite{Sherwood03}, and conditional branch counts \cite{balasubramonian00}. The authors found that the three techniques agree on phases 85\% of the time; BBVs, however, were the most effective at detecting performance changes, finding stable phases, and providing higher sensitivity, i.e., higher probability that the phases are accurately predicted \cite{grunwald98}. The authors also found that classifying phases using instruction working sets identified phases that were 30\% longer than using BBVs. Gu et al. \cite{gu06} presented an overview of various phase classification techniques and metrics, such as the minimum possible size of detected phases, online/offline phase classification, and types of application characteristics used for the classification.  

In this paper, we discuss recent phase classification techniques, focusing on the advances since the techniques discussed in \cite{dhodapkar03} and \cite{gu06}. We survey the phase classification techniques with respect to the characteristics they analyze, the types of applications for which the techniques are effective, when phases are classified, etc. We specifically emphasize discussions of new phase classification techniques for multithreaded applications, which have become more mainstream in the past few years. Finally, we explore the gaps in the state-of-the-art, and propose future research directions for addressing those gaps.    

The remainder of the paper is organized as follows: in Section \ref{uses}, we consider some important uses of phase classification. In Section \ref{back}, we present a brief overview of phase classification techniques presented prior to 2006. In Section \ref{sec:taxonomy} we present a taxonomy of phase classification, and discuss some defining characteristics of phase classification techniques in Section \ref{char}. We discuss online and offline phase classifications techniques in Sections \ref{offline} and \ref{online}, respectively. Within each of these sections, we separate the techniques into those developed for serial and parallel applications. Even though phase classification has been widely studied, there are still some major gaps in the existent techniques, with respect to multithreaded applications, big data applications, Internet of Things computing paradigms, etc. Thus, in Section \ref{future}, we discuss some of these current gaps and future research directions to address them.

\section{Motivations for Phase Classification}
\label{uses}

Several optimization techniques rely on accurate phase classification to detect changes in application characteristics. Much research has shown that leveraging phase characteristics enables a finer grained optimization potential by specializing the system configurations to different execution phases. To motivate this survey, we briefly discuss some of the optimizations that use phase classification, including adaptable hardware, thread-to-core assignment, sampled simulations, and hotspot temperature analysis.

\subsection{Hardware Resource Adaptability}
\label{sysOpt}
Phase-based optimizations use adaptable hardware, such as the adaptable cache presented by Zhang et al. \cite{zhang03}, to specialize hardware to phases' resource requirements without incurring performance overhead \cite{dhodapkar02} \cite{huang03} \cite{Sherwood03} \cite{chen07} \cite{gordon08}. Adaptable hardware allows specialized hardware configurations (e.g., cache associativity, capacity, and line size \cite{zhang03}; issue width \cite{folegnani01}; processor voltage and frequency \cite{isci06} \cite{rao04}; instruction windows \cite{buyuktosunoglu00}; and global history register length \cite{juan98}) for different application phases. Phase-based optimization techniques use phase classification to determine the best points at which hardware configurations must be changed in order to best satisfy application needs.

Adegbija et al. \cite{adegbija12} presented a technique that used adaptable caches and phase-based cache optimization to achieve an average energy delay product (EDP) savings of 26\%. The authors developed a phase distance mapping (PDM) approach that mapped the difference in characteristics between a new phase and a \textit{base phase} to the configuration space, in order to determine the new phase's best configuration. Key to the effectiveness of PDM for EDP optimization was the phase classification stage. Similarly, Meng et al. \cite{meng08} used an adaptable cache to reduce processor power consumption. The authors implemented a power manager that monitored the processor's overall power usage to dictate configuration changes.

Although cache optimization is a common focus of phase-based optimization techniques, other adaptable components benefit from phase-based optimization. For example, Dynamic Voltage and Frequency Scaling (DVFS) \cite{isci06} \cite{rao04} \cite{tee11} \cite{bautista08} \cite{burd00} is commonly used to adapt the clock frequency/voltage to variable runtime execution needs. Other components/configurations that benefit from phase-based optimization include the issue queue \cite{ponomarev01,buyuktosunoglu01}, reorder buffer \cite{abella03,dropsho02}, pipeline \cite{efthymiou03,tiwari07}, register files \cite{abella03,dropsho02}, etc.

\subsection{Thread-to-Core Assignment}
\label{t2c}

Thread-to-core assignments offer another means of optimization. Given a heterogeneous-core system, the effectiveness of specialization using heterogeneity is predicated on the scheduling of execution threads to the core that most closely satisfies the thread's execution requirement. By assigning application threads/tasks to different cores based on phases, fine-grained energy or temperature optimization can be achieved \cite{rodrigues13} \cite{khan10} \cite{srinivasan16}. 

In multicore systems, phase changes indicate when the ideal thread-to-core assignments might change. Phase-based thread-to-core assignment policies consider each phase's resource requirements and migrate threads to the cores that most closely match those requirements. For example, Kumar et al. \cite{kumar03}  \cite{kumar04} used dynamic phase mapping on heterogeneous multicore systems to reduce the overall energy consumption of a system by 39\% on average and achieved a maximum of 31\% speedup over a static policy. Their techniques ensured that the threads stayed mapped to the optimal cores throughout application execution despite changing execution requirements.

\subsection{Rapid Design Evaluation and Sampled Simulation}
\label{simulation}
During system design (e.g., in computer architecture), the design must be extensively evaluated to determine the system's functionalities and efficiency. Evaluation is typically initially performed through simulations. In several cases, however, simulating entire benchmarks is unfeasible due to prohibitive simulation times, especially when using cycle-accurate simulators. For example, SimpleScalar \cite{burger97}, a commonly cycle-accurate simulator, is capable of executing 400 million instructions per hour. SPEC benchmarks \cite{limaye18}, many of which execute well over 300 billion instructions would take about a month to run using Simplescalar (or other similar simulators, such as GEM5 \cite{binkert11}). 

Since application phases are typically repetitive throughout the application's execution \cite{sherwood01} \cite{sherwood02}, each distinct phase only needs to be simulated once to estimate the application's overall behavior. Thus, \textit{phase sampling}, predicated by accurate phase classification, can be used to substantially reduce simulation time as compared to running full applications. Sherwood et al. \cite{sherwood02} used random linear projection followed by k-means clustering \cite{macqueen67} to group phases with similar characteristics. They found that they were able to accurately represent the entire application's execution by simulating a single section of each cluster. Their phase classification technique, called SimPoint \cite{sherwood02} \cite{hamerly05}, is commonly used for sampled simulation \cite{patil04} \cite{perelman06} \cite{ardestani12} \cite{heo03} \cite{kumar03} \cite{kumar04} \cite{lau04} \cite{lau05} \cite{meng08} \cite{Najaf09}.

\subsection{Hotspot Temperature Analysis}
\label{design}

An accurate estimate of the highest possible hotspot temperature on a chip can help circuit designers to reliably verify new circuits and accurately estimate lifetime degradation of chips before the circuits reach market \cite{brooks01}. Srinivasan et al \cite{Srinivasan11} used phases to test system limits. By rearranging application phases, they found that they were able to estimate worst-case hotspot temperatures.  

\section{Pre-2006 Phase Classification}
\label{back}

To provide a background for the more recent phase classification techniques, this section presents an overview of the popular phase classification techniques presented before 2006. Many modern techniques are based on these older techniques or use these older techniques as a baseline for testing modern techniques \cite{perelman06} \cite{peleg07} \cite{shen07} \cite{ratanaworabhan08} \cite{khan10} \cite{khan11}  \cite{sembrant_11}  \cite{zhang15} \cite{bui17}. We direct the reader to \cite{dhodapkar03} for a survey that details these pre-2006 techniques.

\subsection{Basic Block Vectors}
\label{bbv}
Sherwood et al. \cite{sherwood02} \cite{Sherwood03} presented a phase classification technique that uses basic blocks as a microarchitecture independent way to identify phase changes. To implement this classification technique, the frequency information for each basic block---a block of code with one entry and one exit---is stored in Basic Block Vectors (BBVs). The BBVs of different instruction intervals are then compared (e.g., using Euclidean distance) to determine the similarity between the different intervals. Similar intervals are thereafter clustered to form phases. Dhodapkar et al. \cite{dhodapkar03} determined that BBV techniques provide more stable phases and higher sensitivity than other phase-classification techniques. However, BBVs have since been shown to be inaccurate when applications have a large amount of last level cache (LLC) misses \cite{ annavaram04} \cite{carlson13} \cite{huffmire06}.

\subsection{Instruction Working Sets}
\label{iws}
Dhodapkar et al. \cite{dhodapkar02} presented a phase classification technique that utilizes the Instruction Working Set for characterizing the different phases. Instruction working sets are application instructions that are executed at least once during application execution. The authors used working set signatures, a lossy-compressed working set representation. In a working set signature, a hash function is used to map working set elements into vectors. The authors recorded working set signature vectors every 100,000 instructions. The proposed technique featured a user-defined threshold value to determine the sensitivity of phase classification. A larger threshold value means that a greater change in application characteristics must occur for a phase change to be detected. As such, there will be fewer phase changes detected, and phase-based optimizations will occur less frequently.

The authors compared the Manhattan distance, or the sum of differences of each element, between the working set signature vectors after every interval. If the Manhattan distance was below the threshold value, the intervals were considered to be similar and belonging to the same phase. Otherwise, when two intervals' working set signatures differed beyond the threshold, the intervals were considered to belong to different phases for which the system resources must be specialized to optimize execution.
\vspace{-5pt}
\subsection{Conditional Branch Counts}
\label{cbc}

Balasubramonian et al. \cite{balasubramonian00} proposed using conditional branches to determine phase changes. They determined phases by the number of conditional branches executed over an interval. A significant change in the number of branches executed between two intervals indicates a phase change. Unlike most phase classification techniques, the authors did not use a fixed threshold value to determine a significant difference between phases. Rather, the threshold value varies dynamically as the application executes. The conditional branch counter technique and the BBV technique detect a similar percentage of phase changes---phase changes detected vs. total phase changes. However, Dhodapkar et al. \cite{dhodapkar03} found the conditional branch counter technique to be less effective at detecting major phase changes, i.e., phase changes during which application characteristics change significantly.

\begin{figure}[t]
\centering
\includegraphics[width=.3\textwidth]{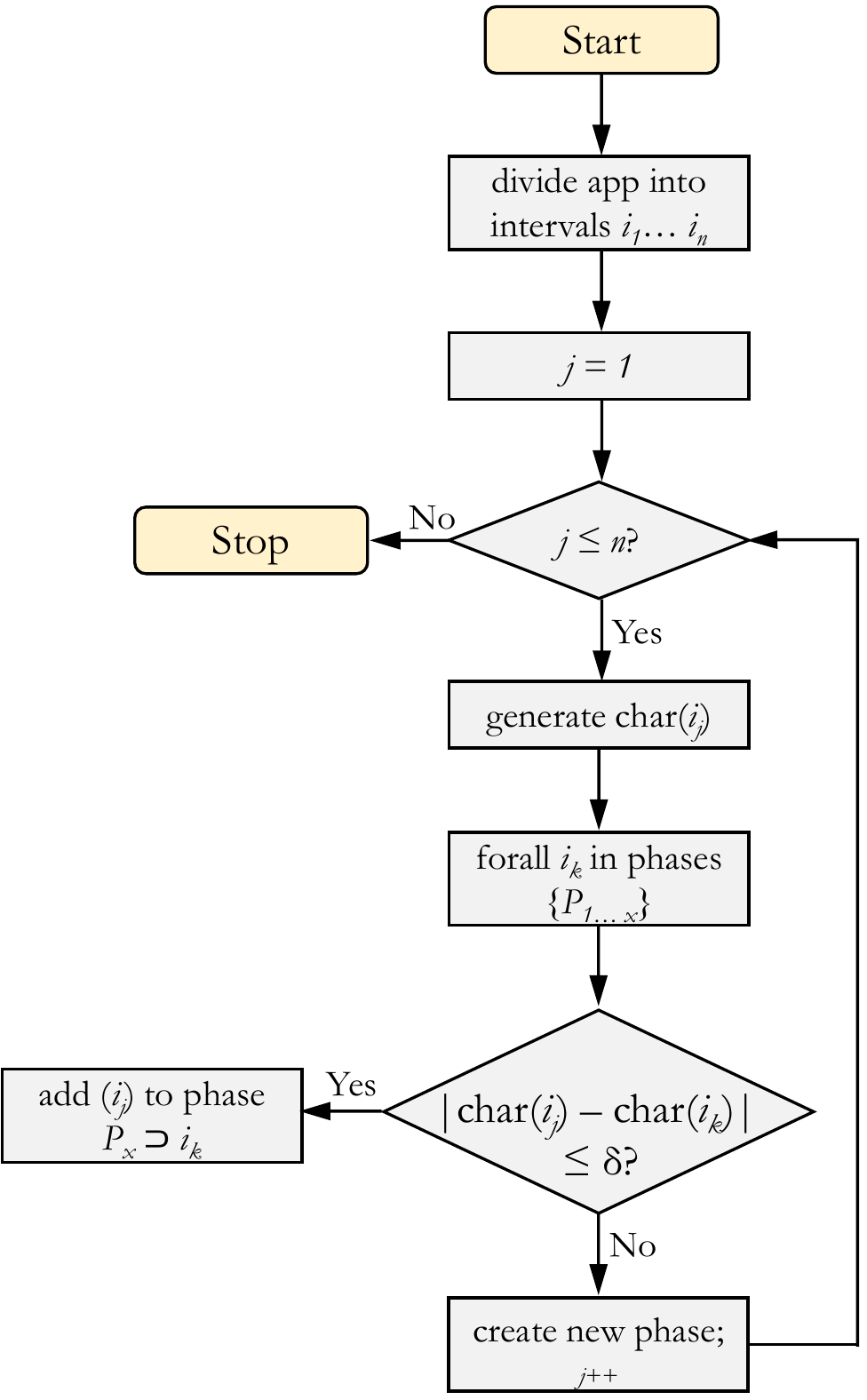}
\caption{High level overview of the phase classification process}
\label{fig:phase_classification}
\end{figure}

\section{Taxonomy of Phase Classification} \label{sec:taxonomy}
\begin{figure*}[t]
\centering
\includegraphics[width=\textwidth]{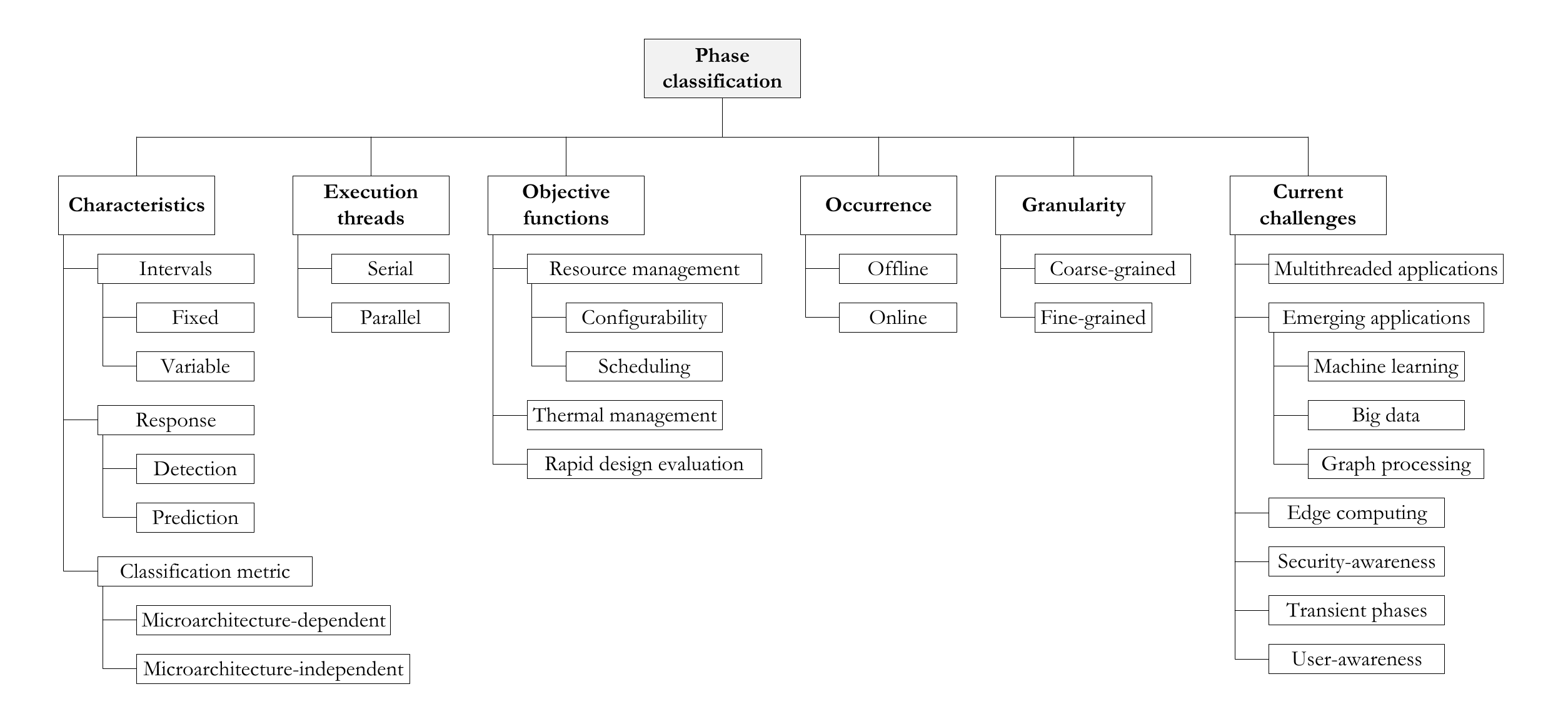}
\caption{Taxonomy of phase classification}
\label{fig:taxonomy}
\end{figure*}

Fig. \ref{fig:taxonomy} depicts our taxonomy of phase classification techniques. Given the the state-of-the-art in phase classification, our taxonomy is based on several factors, including classification parameters, number of threads handled, the target objective functions, when the classification occurs, and the granularity of classification. The taxonomy also includes existent challenges that still remain to be addressed. These different factors and challenges are described in the rest of the paper.

\section{Characteristics of Phase Classification}
\label{char}
Most phase classification techniques follow the same general procedure, depicted in Figure \ref{fig:phase_classification}. First, the application is divided into instruction \cite{huffmire06} \cite{perelman06} \cite{shen07} \cite{peleg07} \cite{khan11} \cite{sembrant_11} \cite{rodrigues13} \cite{zhang15} \cite{srinivasan16} or time \cite{shen_07} \cite{Srinivasan11} \cite{srinivasan12} \cite{ganeshpure13} intervals. For each interval $i$, the application characteristics (\textit{char(i)}) are generated, for example, via design-time simulations or at runtime using statistics from hardware performance counters \cite{casas10} \cite{furlinger08} \cite{gu07} \cite{gu08} \cite{peleg07} \cite{perelman06} \cite{rodrigues13} \cite{shen_07}. Application characteristics that can be used include basic block vectors \cite{sherwood99,sherwood01}, memory accesses \cite{huffmire06} \cite{shen07} \cite{gu08}, power consumption \cite{Srinivasan11}, stalls \cite{srinivasan16}, etc. 

The different intervals' application characteristics are then analyzed to determine intervals that differ by less than or equal to a threshold $\delta$. Although the analysis techniques vary for different phase classification techniques, a sample technique is to use the Manhattan distance between two intervals' application characteristics \cite{sherwood02} \cite{chetsa13} \cite{khan11} \cite{peleg07} \cite{ratanaworabhan08} \cite{sembrant_11} \cite{bui17} \cite{srinivasan12} \cite{zhang15}. Intervals with similar characteristics are then clustered to form phases. When new intervals are encountered, the intervals are added to existent phases or used to form a new phase, depending on the disparity between the new interval's and previously encountered intervals' characteristics. 

Different factors and characteristics come into play when choosing a phase classification technique, or when designing a new one. As such, it is important for users to understand the different possible phase classification characteristics, and how these characteristics may affect phase classification accuracy. In this section, we discuss three defining phase classification characteristics: \textit{intervals}, \textit{classification metrics}, and \textit{phase detection vs. prediction}.

\subsection{Intervals}
\label{intervals}
An interval is the length of time or number of instructions that forms the granularity with which the application characteristics are profiled. Intervals that display similar characteristics are clustered together to form phases. Intervals vary in length depending on the application and the phase classification technique. The granularity, or interval length, of a phase classification technique is most commonly measured in number of instructions. 

One way to categorize intervals is with respect to their granularity. \textit{Fine-grained intervals} are shorter (e.g., on the order of hundreds or thousands of instructions) and require that application characteristics be recorded more frequently. As such, application characteristics represent smaller sections of code, and smaller phases can be detected more accurately \cite{bui17}. However, this enhanced detection comes at the cost of increased overhead---phase classification time and storage space for each interval's information \cite{huang03} \cite{magklis03}. \textit{Coarse-grained intervals}, on the other hand, are longer and enable faster phase classification, since fewer intervals need to be compared. However, coarse-grained intervals may obscure smaller phases, making them more difficult to detect \cite{bui17}. Therefore, the interval granularity must be chosen to balance both speed and classification accuracy. 

Intervals can alternatively be categorized in terms of the variability of their lengths as \textit{fixed length} or \textit{variable length} intervals. Fixed length intervals \cite{huffmire06} \cite{Srinivasan11} \cite{khan11} \cite{srinivasan16} \cite{bui17} \cite{khan11} \cite{shen_07} \cite{rodrigues13} \cite{shen07} \cite{srinivasan12}, as the name implies, remain static throughout phase classification. As a result, they are simpler to implement and are more commonly used. However, because the interval size does not change during phase classification, fixed-length intervals can lead to inaccurate phase classification if the phase behavior changes at a different periodicity than the intervals. Lau et al. \cite{lau_05} found that fixed-length intervals can become mismatched with the naturally-periodic behavior of applications. This mismatch results in phase changes occurring in the middle of an interval instead of near the edge, thus reducing the accuracy of phase classification. 

Variable length intervals \cite{peleg07} \cite{gu08} \cite{chiu16} \cite{zhang15}, on the other hand, allow the intervals to closely match the periodicity of the different phases and enable more accurate phase classification. However, variable length intervals are more complex, since they may require more detailed application analysis at design time or an online algorithm with a feedback loop for runtime phase classification \cite{gordon-ross07}. 

\subsection{Classification Metrics}

Classification metrics refer to the application characteristics that are used to determine the similarity or variability between intervals. Classification metrics can be broadly categorized into two: \textit{microarchitecture-dependent} and \textit{microarchitecture-independent} metrics \cite{phansalkar05,phansalkar07}. Microarchitecture-dependent metrics, as the name implies, are application characteristics that are a function of the system microarchitecture on which the application is run. These metrics are typically obtained using hardware performance counters or simulators, and would change if the application is run on a different microarchitecture or configuration. Examples of microarchitecture-dependent metrics include cache miss rates, instructions per cycle (IPC), translation look-aside buffer (TLB) miss rates, branch misprediction rates, etc. These characteristics depend on specific hardware structures (e.g., cache, branch predictor) and would change if different architectures or configurations are used. 

Microarchitecture-dependent metrics are straightforward to obtain, given that simulators and hardware performance counters are ubiquitous in modern-day computer systems. In addition, microarchitecture-dependent metrics make phase classification techniques that use them more amenable to being used during runtime. Furthermore, these metrics may better detect dynamic changes in application behavior, such as new data inputs or a previously unknown stimuli. However, a major downside of these metrics is that they can hide underlying inherent program behavior, potentially leading to inaccurate phase classification \cite{hoste07}. It is possible for two applications to exhibit similar microarchitecture-dependent characteristics on one system but behave drastically different on another. To prevent this downside, \textit{microarchitecture-independent} metrics \cite{hoste07,eeckhout05,poovey09} use characteristics that depict the inherent characteristics of the applications being characterized. Examples of microarchitecture-independent metrics include basic block vectors, memory access patterns, working set size, branch behavior, etc. These metrics typically need to be collected using binary instrumentation tools, such as ATOM, Pin, Valgrind, and are easier to collect at design time \cite{reddi04,nethercote07,srivastava04}. 

\subsection{Detection vs. Prediction} 
\label{dp}

Phase classification techniques can be divided into two types---\textit{detection} and \textit{prediction}---depending on when the actual classification occurs. Detection techniques are reactive; they compare application execution metrics after they are recorded---i.e. after the interval has ended---to those of the previous interval \cite{casas10} \cite{khan11} \cite{Srinivasan11} \cite{srinivasan12} \cite{ganeshpure13} \cite{chetsa13} \cite{rodrigues13} \cite{brankovic14} \cite{chiu16} \cite{srinivasan16} \cite{bui17} \cite{nagpurkar06_online} \cite{benomar14}. Since the comparison happens after an interval, phase classification occurs after a phase change has occurred \cite{gu06}. 

Conversely, the goal of a prediction technique is to classify a new phase \textit{before} the new phase change occurs \cite{shen07} \cite{gu07} \cite{sembrant_11} \cite{sembrant12} \cite{chiu18} \cite{curtis06} \cite{curtis08}. These techniques feature a training period, which occurs at design time or during the beginning of application execution. During the training period, information (e.g., execution statistics) about the application's variable phase patterns are gathered. After the training period, a phase prediction technique analyzes newly acquired application characteristics and information about the application's typical phase patterns. These characteristics and analyses are then used to predict phase changes. While the analyses used vary between prediction techniques, the general idea is that predictions can be made about when and how an application's execution characteristics will change based on previously recognized and detected patterns. The biggest drawback to phase prediction techniques is that they are most effective with applications that have predictable phases or applications whose phase structure does not differ with new inputs \cite{shen07}. Since detection techniques are significantly more common than prediction techniques, this survey focuses on detection-based phase classification techniques.


\begin{table*}[]
\centering
\caption{Overview of offline phase classification techniques}
\renewcommand{\arraystretch}{1}
\label{table:ApplDomainCompare1}
\resizebox{1\textwidth}{!}{%
{\begin{tabular}{|c|c|c|c|}
\hline
 & \multirow{8}{*}{Serial} & Interval type & {Variable \cite{gu08} \cite{shen07}; fixed \cite{zhang15} \cite{huffmire06} \cite{Srinivasan11} \cite{shen_07}  \cite{brankovic14}}\\ \cline{3-4}
  
 & & Interval width & 10M \cite{zhang15} \cite{brankovic14} and 1M instructions \cite{huffmire06}; 666.6 us \cite{Srinivasan11}; 10ms \cite{shen_07}; context switch \\
 
 & &  &  points \cite{gu08}; reuse distance \cite{shen07} \\ \cline{3-4}
 
 & & Classification & Basic blocks \cite{zhang15} \cite{ratanaworabhan08} \cite{shen_07}; cache misses \cite{huffmire06} \cite{gu08} \cite{shen_07}; power consumption \cite{Srinivasan11};\\
 
 & & metrics & memory accesses \cite{shen07}; instructions per cycle \cite{shen_07}; static/dynamic \\
 
 & &  & instruction ratios \cite{brankovic14} \\ \cline{3-4}

 & & Analysis method & Manhattan distance \cite{zhang15} \cite{brankovic14}; basic block execution frequency \cite{ratanaworabhan08} \cite{shen_07};\\
 
 & &  & digital signal processing \cite{huffmire06} \cite{Srinivasan11} \cite{shen07}; pattern analysis \cite{gu08}; principal component analysis \cite{hoste07}; \\
 & & & genetic algorithm \cite{hoste07}\\ \cline{2-4}
 
  & \multirow{5}{*}{Parallel} & Interval type & Variable \cite{peleg07} \cite{furlinger08}  \\ \cline{3-4}
 
 & & Interval width & Several billion instructions \cite{peleg07} \\\cline{3-4}

 & & Classification & Instruction execution \cite{peleg07}; function execution \cite{furlinger08}; instructions per cycle \cite{casas10} \\
 
 & & metrics & \\ \cline{3-4}
 
 & & Analysis method & Manhattan distance \cite{peleg07}; pattern analysis \cite{furlinger08}; digital signal processing \cite{casas10} \\\hline

\end{tabular}%
}}
\end{table*}


\section{Offline Phase Classification}
\label{offline}

Offline phase classification techniques use execution characteristics, such as basic blocks and execution traces, which are gathered during compile time or through a priori application analysis. These techniques are generally easier to develop than online phase classification techniques, since online techniques have stringent constraints where negative impacts on application execution time must be minimized. In addition, offline techniques need fewer runtime resources for storing application characteristics during classification. Only information about phase IDs and phase change locations need to be stored during runtime.

Table \ref{table:ApplDomainCompare1} summarizes the surveyed offline phase classification techniques, which we categorized as \textit{serial}---for classifying single-threaded applications---and \textit{parallel}---for classifying multi-threaded applications. This section describes different serial and parallel phase classification techniques.

\subsection{Serial Offline  Phase Classification}
\label{serial_off}

Several techniques have been developed to classify phases in single execution threads. These techniques are typically simple to design, since they must only record and analyze application characteristics for one application thread, without needing to coordinate phase information among multiple threads. 

Zhang et al. \cite{zhang15} focused on improving the accuracy of basic block vector techniques for single-threaded applications. The authors examined the execution sequence of fine-grained phases and found that these phases' patterns could be used to predict course-grained phase execution. The authors presented multilevel phase analysis, a technique that combines analyses of different phase granularities. During a training period, their technique identified both fine-grained and coarse-grained phases, and stored a list detailing the sequence of fine-grained phases in each new coarse-grained phase. The authors classified fine-grained phases by comparing the Manhattan distance between basic block vectors. They used an interval size of 10 million instructions while determining fine-grained phases and used outermost loop boundaries to determine coarse-grained phases. After the training period, the classification technique identified $x$ fine-grained phases---the authors found 5 to be sufficient in their experiments. The technique then compared the execution sequence to that of previously-identified course-grained phases. If the technique discovered the same sequence of course-grained phases in the look-up table, it would accurately predict the rest of the fine-grained intervals in the course-grained phase. If the fine-grained phase sequence did not match any sequence in the table, execution continued until a course-grained phase was identified.

Ratanaworabhan et al. \cite{ratanaworabhan08} also used basic blocks to classify single-threaded application phases. The authors used ATOM \cite {srivastava94}, an application analysis tool, to identify basic blocks and assign each basic block a unique identification number. During phase classification, the authors identified a phase change when a significant number of new basic blocks executed in a short period of time. They then created a phase identification signature by storing the basic block identification numbers of the new basic blocks that indicated the phase change. By comparing this signature to future sequences of basic blocks, they were able to identify any repetitions of a phase.

Huffmire et al. \cite{huffmire06} presented a wavelet-based technique that used a fixed-length interval of 1 million instructions and a digital signal processing technique to classify phases in single-threaded applications. Wavelets, commonly used in digital signal and image processing, are mathematical functions that encode both frequency and spatial information, unlike BBVs, which store only frequency data \cite{huffmire06} \cite{sherwood01}. For their phase classification technique, the authors used wavelets to store L1 cache access data. They began by gathering an application trace of memory accesses, and then used k-means clustering to analyze the wavelets. Specifically, the authors were interested in analyzing wavelet signatures of all L1 cache accesses to predict L2 cache misses. By comparing these predictions across subsequent intervals, they were able to detect phases.

Srinivasan et al. \cite{Srinivasan11} also proposed an offline phase classification technique that used fixed-length intervals. However, instead of tracking their intervals by number of instructions like most other techniques, they used an execution time---666.6 us. They also used the simulated power traces within the different intervals to determine phases, based on the observation that spatial power dissipation remains unchanged within a phase. From the power trace, they created a power vector that stores a moving average of 500 power values. They detected the beginning of a phase if three consecutive power vectors were similar, and detected the end of a phase if two consecutive power vectors were dissimilar. Conversely, most other offline phase classification techniques use simulated hardware counters or other frequency-based metrics, such as cache miss rates \cite{huffmire06} \cite{gu08} \cite{shen_07} \cite{gu07} and basic block execution frequency \cite{zhang15} \cite{ratanaworabhan08} \cite{shen_07} for characterizing application phases.  

Traditional interval lengths---on the order of several thousands to millions of instructions or cycles---are not always acceptable for accurate phase classification. Bui et al. \cite{bui17} studied past interval lengths and found that traditionally-sized intervals do not always detect a significant number of application phases. This behavior occurs because the interval sizes are too coarse to detect small phases, and phase based optimization opportunities are missed. The authors found that classifying smaller phases can improve the optimization gained by phase-based optimization techniques. They argued that modern hardware optimizations offset much of the overhead from using fine-grained phases and proposed using 'super fine-grained intervals' to better classify phases in single threaded applications. The 'super fine-grained intervals' they suggested are on the order of tens, hundreds, or thousands of application cycles.

Although there are some phase classification techniques that use factors other than hardware data to determine phase changes---such as Srinivasan et al.'s \cite{Srinivasan11} use of power consumption---many phase classification techniques use some combination of hardware data. Gu et al. \cite{gu08} presented one such technique of hardware data use in applications run by a Java Virtual Machine. To classify phase changes, they analyzed a trace of the number of L1 instruction cache misses. Using pattern analysis, they looked for changes in the density of L1 instruction misses across intervals. Gu et al. aimed to classify variable and course-grained phases.

Shen et al. \cite{shen07} presented another phase classification technique that utilizes hardware counters---specifically, they used a memory access trace and wavelet analysis to identify phase changes during a training period. During the training period, the authors executed several different iterations of an application, using different inputs each time. They used a digital signal processing technique---a Discrete Wavelet Transform---to analyze a memory access trace and find significant changes in the access pattern. Their technique was designed for applications with large, well-defined phases whose phase sequences do not differ significantly with different inputs. More specifically, ideal applications differ only in the number of times sections of the phase sequence are repeated, referred to as "exponents." A well-defined phase is one that exhibits a significant change in performance characteristics from that of the code surrounding it. To predict phases, the authors analyzed new application inputs and phase pattern exponents. However, their technique does not work if many phase pattern exponents did not change in conjunction with an input parameter.

Sometimes, only a limited knowledge of the executing application exists---the phase classification mechanism may not have access to characteristics such as hardware counters. Shen et al. \cite{shen_07} proposed \textit{active profiling}, a phase classification technique that works on applications in binary form and requires no knowledge of loop or function structure, as in several offline phase classification techniques. The authors found that they could detect phases by controlling the application input. By controlling the input---specifically, by issuing a series of nearly identical requests---they were able to get the application to output an artificially regular pattern, or a particular behavior that repeats a prearranged number of times. To execute phase classification, the authors acquired a basic block trace of an application using the repeated request input. In their active profiling technique, the authors selected basic blocks that executed exactly the same number of times as the input was repeated. Next, they verified that the execution of all the basic blocks of one type were evenly spaced during execution. The authors ignored basic blocks with executions that were not evenly spaced, and removed false positives by running the application with a real input. Basic blocks that were not executed the same number of times as there are input requests were identified as false positives. Of the remaining basic blocks, the first instance of each indicated a phase change.  

Thus far, we have discussed phase classification techniques that were designed to work on hardware-only processors or virtual machines. However, hardware/software (HW/SW) co-designed processors behave differently. Brankovic et al. \cite{brankovic14} found that traditional phase-classification techniques are unsuited for HW/SW co-designed processors. The authors analyzed BBVs for use as a phase classification metric and found that each BBV's execution time varies significantly more when run on HW/SW co-designed processors versus hardware-only processors. Therefore, the authors theorized that one cannot assume that phases of applications classified with BBVs will behave as expected when run on a HW/SW co-designed processor. For HW/SW co-designed processors, the authors proposed Transparent Optimization Layer Description Vector Phase Classification (TDV). The Transparent Optimization Layer is the software layer in a HW/SW co-designed processor that dynamically analyzes, profiles, translates, and optimizes instructions from a guest instruction set architecture (ISA) to the host ISA. TDV uses intervals and stores 19 execution statistics that contain information about the static/dynamic instruction ratio and the instructions mix of the application to be executed. The technique then compares the statistics for each interval with the statistics from previous intervals to determine the different phases.

\subsection{Parallel Offline  Phase Classification}
\label{parallel_off}

Parallel phase classification techniques cater to multithreaded or multiprogrammed applications by determining the application phases while taking into account the individual application threads. In general, due to multithreaded applications' intrinsic characteristics, parallel techniques are more complex than serial techniques. Since executing threads may be competing for the same resources, parallel thread execution can affect commonly used performance metrics such as cycles per instruction (CPI) and cache misses, rendering such performance metrics ineffective for phase classification \cite{peleg07}.

In addition, parallel applications can have data-sharing or resource-sharing threads, which complicate phase classification  \cite{sinharoy05} \cite{herescu05}.  Although serial phase-classification techniques are simpler, multithreaded applications are becoming increasingly common in emerging computer systems. Thus, phase classification in multithreaded applications remains an important research area that necessitates novel techniques. 

Peleg et al. \cite{peleg07} addressed multithreaded phase classification by using changes in code execution frequency to classify phases. Since traditionally used hardware counters such as cache misses and CPI can be influenced by all threads in a multithreaded application, the authors used a performance metric that would not be influenced by other threads. The proposed technique sampled instruction addresses to capture the profile of the code execution frequency. The authors then created a histogram of sampled addresses to represent the applications' BBV. 

Furlinger et al. \cite{furlinger08} also used code execution frequency to classify phases. However, the authors recorded connections between functions and manually determined phase start and end locations. They used ompP \cite{furlinger08ompp}, a profiling tool, to record a call graph---the connections between different functions---of the executing application. For accurate phase classification, the authors modified ompP to track all predecessor nodes in the call graph, and the number of times each predecessor node was executed. The authors defined a predecessor node as a parent node or a sibling node of the same level as the current node. They then manually analyzed the call graph to detect patterns in the execution. Their phase classification technique required the designer to determine which nodes classify as a pattern, and thus could not be automated.

Casas et at. \cite{casas10} used a signal processing technique---a discrete wavelet transform---to classify phases in multithreaded applications. They found that, by using a signal processing technique on data from a trace file, they could detect only the most relevant frequencies of an application’s execution---the frequencies most strongly related to the loops of an application’s source code. By using a discrete wavelet transform, they were able to acquire information about the frequency and the locations of the signal. They found several different signals to be suitable for phase classification---the number of processes computing, the total duration of computing bursts, the number of point-to-point MPI calls, the number of collective MPI calls, the instructions per cylce, and the number of outstanding messages at a given point.

\section{Online Phase Classification}
\label{online}

Although offline phase classification has fewer design constraints than online classification, it is often impractical. Offline techniques require a priori knowledge of the executing applications, and many systems, such as general purpose systems (e.g., smartphones and tablets), may not have complete design time application knowledge. In addition, runtime application variability, such as new data inputs, may cause the application to behave differently. Online phase classification addresses these drawbacks by classifying the phases during runtime. However, runtime applications must execute quickly to avoid adversely impacting the executing application's latency. In addition, online phase classification techniques must be reliable and accurately classify phases with minimal overheads (e.g., energy, storage, etc.). Table \ref{table:ApplDomainCompare2} summarizes the serial and parallel online phase classification techniques described in this section.

\begin{table*}[h]
\centering
\caption{Overview of online phase classification techniques}
\renewcommand{\arraystretch}{1.2}
\label{table:ApplDomainCompare2}
\resizebox{1\textwidth}{!}{%
{\begin{tabular}{|c|c|c|c|}
\hline

& \multirow{6}{*}{Serial} & Interval type & Fixed \cite{srinivasan16} \cite{chetsa13} \cite{gu07} \cite{khan11} \cite{sembrant_11}; variable \cite{chiu16} \\ \cline{3-4}
 
 & & Interval width & 10K-15K instructions \cite{srinivasan16}; 10M \cite{khan11}; 100M \cite{sembrant_11} instructions; 1s \cite{chetsa13}; 100K branches \cite{chiu16} \\ \cline{3-4}

 & & Classification & Resource bottlenecks \cite{srinivasan16}; execution vectors \cite{chetsa13}; cache misses \cite{gu07}; \\
 
 & & metrics &  instruction type vectors \cite{khan11}; branch edges/method entry/returns \cite{chiu16}; \\
 
 & & & conditional branches \cite{sembrant_11} \\\cline{3-4}
 
 & & Analysis method & Manhattan distance \cite{srinivasan16} \cite{chetsa13} \cite{gu07} \cite{sembrant_11}; digital signal processing \cite{khan11} \cite{chiu16} \\\cline{2-4}

 & \multirow{5}{*}{Parallel} & Interval type & Fixed \cite{rodrigues13} \cite{sembrant12} \cite{ganeshpure13} \\\cline{3-4}
 
 & & Interval width & 50K - 200K instructions  \cite{rodrigues13}; 100M instructions \cite{sembrant12}; execution time \cite{ganeshpure13} \\\cline{3-4}

 & & Classification & Instruction type vector \cite{rodrigues13}; conditional branch execution \cite{sembrant12}; task execution \\
 
 & & metrics & frequency \cite{ganeshpure13} \\\cline{3-4}

 & & Analysis method & Manhattan distance \cite{rodrigues13} \cite{ganeshpure13}; Clustering algorithm \cite{sembrant12} \\\hline

\end{tabular}%
}}
\end{table*}

\subsection{Serial Online Phase Classification}
\label{serial_on}

Hardware characteristics are often used to classify phases in online techniques. Unlike microarchitecture-independent characteristics gathered from traces in offline techniques, online techniques gather hardware characteristics directly from system hardware counters during execution. Srinivasan et al. \cite{srinivasan16} used hardware characteristics to track the frequency of resource bottlenecks experienced by applications. Specifically, they recorded cache stalls, branch mispredicts, IPC, and resource stalls---the number of cycles the instruction dispatch is stalled due to a blocked instruction queue, re-order buffer, or issue width. They stored this information in a vector referred to as a Bottleneck Type Vector (BTV). As with BBVs, the Manhattan distance between BTVs were compared to determine which intervals behaved similarly to form phases.

Chesta et al. \cite{chetsa13} used hardware counters, network bits sent/received, and disk read/write counts to classify phase changes. The authors used hardware counters to record the number of retired instructions, last level cache references and misses, and branch instructions and misses. The proposed technique utilized fixed-length intervals with a sampling frequency of one second. System characteristics were stored in execution vectors and phase changes were determined using the Manhattan distance between two execution vectors. When the distance exceeded a predefined threshold, the phase was considered to have changed.

Gu et al. \cite{gu07} used microarchitecture-level hardware events to classify and predict longer-than-average phases of different lengths. During the training period for their prediction phase classification technique, the authors recorded the density of L1 instruction cache (i-cache) misses. They then computed the difference of L1 i-cache misses between two intervals. If the difference was more significant---exceeding a higher threshold---than that of the previous two intervals, a new pattern was started. The prediction used a table-based technique that stored information gathered during the training period in a table. The technique then referenced the table during phase prediction. The authors stored identified patterns in a pattern database, or table. Along with the patterns, the table also stored three common \textit{repetition distances}. A repetition distance is the number of instructions between two occurrences of the same phase. Analyzing the repetition distance enabled the proposed technique to predict phase changes before they occurred.

Rather than using hardware characteristics for phase classification, Khan et al. \cite{khan11} presented a classification technique that counted executed instruction types and stored the information in an \textit{Instruction Type Vector (ITV)}. The authors considered various instruction types: integer ALU, complex, branch, load, store, floating point ALU, multiply, and divide instructions. Their technique utilized fixed-length intervals of 10 million instructions, and counted occurrences of the instruction types within each interval. The authors then used the Manhattan distance between two intervals to cluster the intervals into application phases. One key feature of this technique is that it uses microarchitecture-independent application characteristics for phase classification. However, the technique may accrue additional computation and storage overhead, since the ITV must be computed and the intermediate data must be stored during runtime.

Chiu et al. \cite{chiu16} found that they could use variable-length intervals to improve classification accuracy. The authors proposed a technique that first traced an application's execution, profiling the function calls/returns and conditional branch instructions every 100,000 branches. Then, the technique applied a Gaussian mixture model \cite{mclachlan04} to cluster the different intervals based on the number of executed branches. If a cluster's number of executed branches was different from previous clusters, the cluster was determined to belong to a different phase. The major drawback to this technique is that it requires an offline training period, and cannot be fully implemented as a runtime technique.

Sembrant et al. \cite{sembrant_11} presented work to reduce the overhead of BBVs. They suggested using Precise Event Based Sampling (PEBS) \cite{intel10} to directly measure sparse BBVs, or randomly sampled BBV execution frequencies. PEBS, developed by Intel, tracks the number of events that have occurred and records the CPU's state after $N$ events, where $N$ is a value chosen by the user. By declaring a perf\_event to record the address of of every Nth branch instruction, the authors were able to record basic block frequency vectors. The authors found that counting only conditional branches sufficed for achieving accurate BBV identification. In addition, the authors developed a C/C++ library, called ScarPhase, that consolidates their phase-classification techniques for easy use.
\subsection{Parallel Online Phase Classification}
\label{parallel_on}

Some phase classification techniques modify a more simple serial and/or offline technique for online use with parallel applications. For example, Rodrigues et al. \cite{rodrigues13} modified the serial technique presented by Khan et al. \cite{khan11}. The proposed technique reduced the number of entries in the Instruction Type Vector (ITV) to four---INT (integer), FP (floating point), iBJ (branch), and Mem (load and store). The authors then used the Manhattan distance between ITVs to detect phases. The authors found that they could combine phase classification with dynamic core morphing \cite{rodrigues11} to significantly improve the performance/watt of most multithreaded workloads.

Sembrant et al. \cite{sembrant12} analyzed the phase size in various applications and found that data-parallel applications have shorter phase sizes as the number of threads increases. Because smaller phases indicate that hardware/software changes occur more frequently, runtime phase-guided optimizations often become more costly as the number of threads increases. This makes runtime phase optimization more difficult in highly parallel applications. In addition to their analysis, the authors presented a prediction phase classification technique. Their technique used ScarPhase \cite{sembrant_11} to classify application behavior. To accomplish phase classification, The authors \cite{sembrant12} extended the ScarPhase library to support parallel applications by including the functionality to alternate between threads. ScarPhase classifies a phase for a particular thread and then classifies the phase for whichever thread finishes an interval next.

Ganeshpure et al. \cite{ganeshpure13} designed a phase classification technique to run on an multi-processor system on chip (MPSoC). An MPSoC is a system with multiple processor cores (processing elements---PEs) connected by a Network on Chip. The authors' phase classification technique required one processing element to act as a leader and the rest as followers. The leader could also act as a follower by sharing resources with thread execution. In the classification technique, the followers detected local phases independently and transferred the information to the leader. Each follower stored a Follower Phase Vector (FPV), which is updated every interval. The FPVs stored execution clock cycles for task execution and destination PEs as well as the number of times the task information is encountered. When a new FPV is generated, it is compared to the previous FPV. If the Manhattan distance between FPVs was below a certain threshold, the FPVs were considered matching. A local phase was detected if three consecutive FPV pairs match. If all of the PEs detected local phases before one of the local phases ended, the system detected a global phase. These global phases were then used for thread scheduling.

\section{Future Research Directions}
\label{future}

Even though, as illustrated in this survey, phase classification is a relatively mature research area, there are still gaps in the state-of-the-art that have not yet been fully addressed. This section briefly highlights some of the existing challenges that must be addressed---and areas of computing in which the challenges exist---to fully leverage the benefits of emerging adaptable systems.

\noindent\textbf{Phase classification for emerging multithreaded applications:} One of the most important gaps is the fact that the majority of current effective phase classification techniques are designed for single-threaded applications. However, multithreaded applications are now ubiquitous in modern computer systems, including resource-constrained computing systems. Multithreaded applications are becoming even more important with the emergence of new classes of big data applications, such as machine learning, graph processing, image processing, and databases. As prior work has shown, adaptability will be a necessary feature for computing architectures that target these application domains \cite{doppa17}.

A distinguishing feature of these applications is that they are typically massively parallel with threads that run on large-scale architectures, such as many-core architectures with tens, hundreds, or even thousands of threads \cite{ben18}. In addition, several of these threads may need to interact with each other to enable functional correctness of the executing applications. As such, to enable optimal performance of the architectures for these applications, novel phase classification techniques must be developed for a wide variety of multithreaded applications, from small-scale applications to large-scale massively parallel multithreaded applications. The few currently existing phase classification techniques for parallel applications typically assume independent application threads without explicit consideration of data sharing. There are currently no known phase classification techniques designed to explicitly cater to multithreaded applications that exhibit data sharing.

Multithreaded data-sharing applications, especially, pose significant new challenges for phase classification. Accurately classifying multithreaded applications' phases is much more difficult due to shared resources, inter-core dependencies, and shared data. Data sharing cores may share working sets; a thread's runtime characteristics may depend on the characteristics of another thread running on another core. These kinds of data sharing multithreaded applications are expected to remain a prominent feature of emerging embedded systems. Currently, there is a critical knowledge gap on the implications of inter-core dependencies and data sharing for runtime phase classification in multithreaded applications.

\noindent\textbf{Embedded, mobile, and edge computing:} Despite the prevalence of embedded systems and mobile computing, there are currently no phase classification techniques specifically designed for mobile devices. Previous work \cite{rao17} suggests that the phases of mobile applications may be different from those of traditional desktop application, but a comprehensive study of mobile applications' phases still remains elusive. While existent techniques may be applicable to resource-constrained embedded systems, most phase classification techniques introduce hardware, runtime, or energy overheads. These overheads may be prohibitive for embedded systems with stringent resource constraints. There is still much room for improvement in minimizing the overheads imposed especially by runtime phase classification.

An emerging area of computing that is amenable to adaptability is \textit{edge computing} \cite{varghese16}. Edge computing has emerged as a paradigm, in the framework of the Internet of Things (IoT), where computation is moved closer to edge data-gathering devices in order to mitigate the bandwidth and latency overheads of transmitting data to the cloud for computation. An important characteristic of edge computing systems is the ability of the systems to be adaptable, not only to applications' changing requirements, but also to execution contexts, environmental factors, etc. \cite{dangelo18}. Edge computing design choices like computation migration \cite{shahhosseini19}, for example, rely on the classification of application phases or tasks that will run on the edge device vs. the cloud or fog level of the IoT hierarchy. To fully satisfy the adaptability requirements of edge computing, new phase classification techniques must be developed to specifically satisfy the requirements of edge computing systems, including context awareness, low energy consumption, the need for computation migration, etc.

\noindent\textbf{Security-aware phase classification:} Phase classification may also come into play in security applications. Due to the fact that computer systems operate in dynamic environments, security mechanisms must also be adaptable to the inherent dynamism of the computing environments \cite{portilla10,almorsy14}. Furthermore, different application phases may have different threat levels, and the phases must be classified in order to enable the design of security mechanisms that guarantee the required levels of protection for the different phases. As such, security-aware phase classification techniques need to be developed to incorporate security objective functions as part of the metrics for evaluating the characteristics of the different phases. 

\noindent\textbf{Transient phases:} Another area that warrants further study in phase classification is the impact of \textit{transition phases} \cite{lau05-transition} on the overall classification accuracy. Transition phases refer to the execution periods between stable phases. Most current techniques ignore these transition phases, which, if considered, may substantially change the classification technique's accuracy. Conversely, transition phases, if accurately detected and characterized, may offer additional opportunities for improving the specialization of system resources to application requirements.

\noindent\textbf{User-aware phase classification:} Finally, most current phase classification techniques are monolithic. Even though the main goal of phase classification is to exploit the variety of application execution characteristics for system adaptability, several factors can impact the application behavior during runtime. An application's phase characteristics may change drastically throughout execution as a result of multiple factors, such as new data inputs, system execution conditions, or user requirements. The design of phase classification techniques is currently disjoint with these factors, which can limit the achievable optimization from dynamically adaptable computing. Thus, new dynamic phase classification techniques are required to robustly handle and integrate known application and system information with predictive models for runtime application and system behavior changes, and variable user requirements, which may be unknown at design time.
\vspace{10pt}
\section{Conclusions}
\label{conclusion}
The benefits derived from adaptability are directly tied to the accuracy of identifying the points at which the system configurations must be changed. Thus, phase classification is an important initial step in the design of adaptable computer systems that can be specialized to variable application requirements. Phase classification also offers other benefits, including speeding up research simulations, enabling efficient runtime thread-to-core assignments, etc. 

In this paper, we presented a survey of phase classification techniques for identifying program phases. We categorized the different techniques based on several important characteristics, in order to highlight the techniques' similarities and differences. We also highlighted some of the gaps in the state-of-the-art to expose future important research directions on phase classification. We hope that this survey will provide researchers with valuable insights into the state-of-the-art in phase classification, and direction on how to further enhance the benefits of adaptable computer systems for a wide variety of emerging applications. 

\bibliographystyle{abbrv}
\bibliography{bibItems}

\begin{IEEEbiography}[{\includegraphics[width=1in,height=1.25in,clip,keepaspectratio]{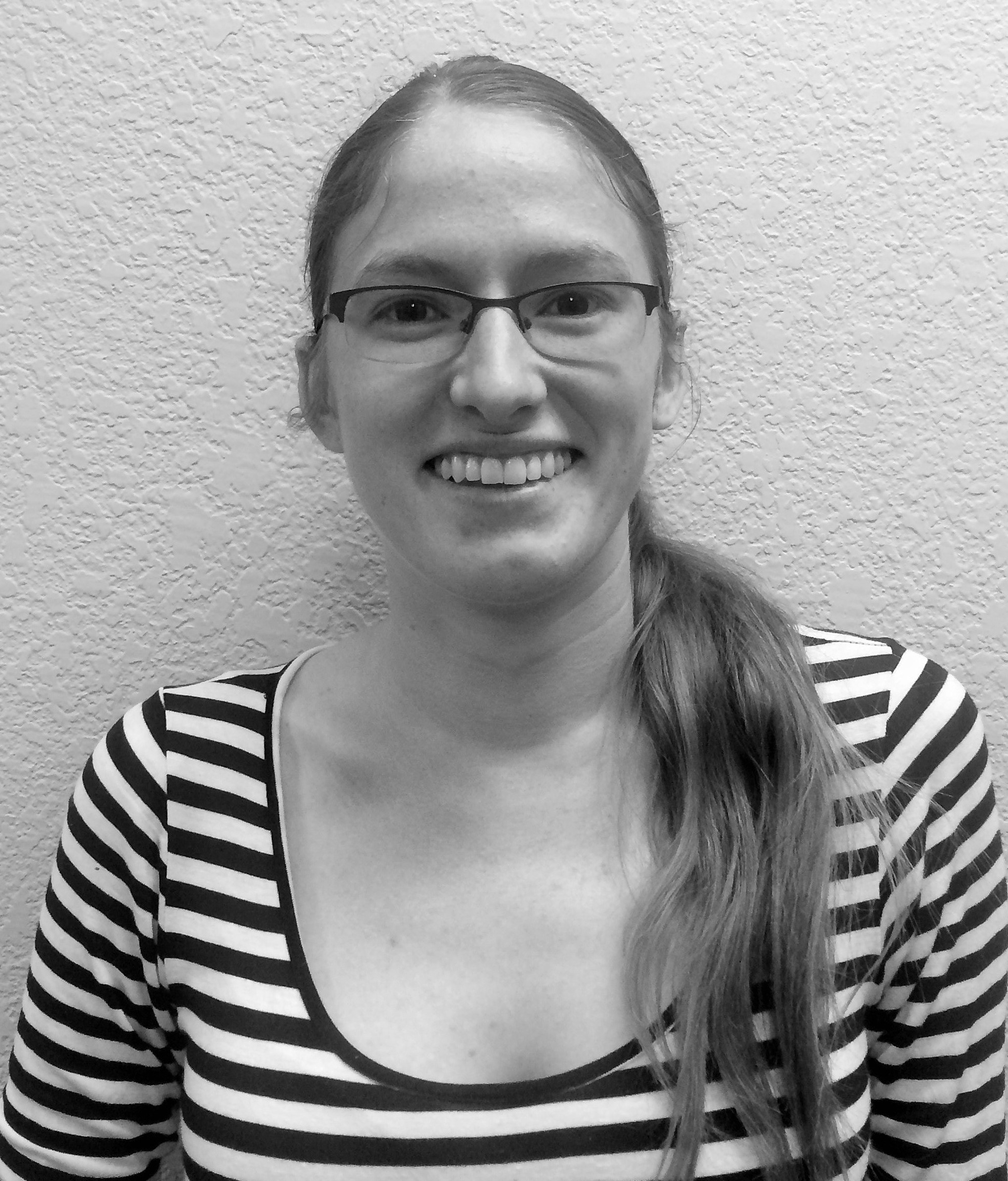}}]{Keeley Criswell} received her M.S. in Electrical and Computer Engineering from the University of Arizona in 2018 and her B.S. in Physics and Computer Science from Thiel College in 2015.
\end{IEEEbiography}

\begin{IEEEbiography}[{\includegraphics[width=1in,height=1.25in,clip,keepaspectratio]{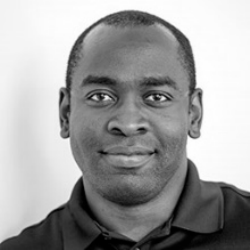}}]{Tosiron Adegbija} (M'11) received his M.S and Ph.D in Electrical and Computer Engineering from the University of Florida in 2011 and 2015, respectively and his B.Eng in Electrical Engineering from the University of Ilorin, Nigeria in 2005. 

He is currently an Assistant Professor of Electrical and Computer Engineering at the University of Arizona, USA. His research interests are in computer architecture, with emphasis on adaptable computing, low-power embedded systems design and optimization methodologies, and microprocessor optimizations for the Internet of Things (IoT). 

Dr. Adegbija was a recipient of the CAREER Award from the National Science Foundation in 2019 and the Best Paper Award at the Ph.D forum of IEEE Computer Society Annual Symposium on VLSI (ISVLSI) in 2014.
\end{IEEEbiography}

\balance

\end{document}